%% file: thesis.tex
\documentclass[12pt,oneside,letterpaper,chapterprefix=on,numbers=noenddot]{scrbook}
\usepackage{algorithm}
\usepackage{subfig}
\usepackage{pdfpages}

\usepackage[margin=1in,includefoot,heightrounded]{geometry} 	
\usepackage[english]{babel}  								
\usepackage{amsmath}									

\usepackage{indentfirst} 									
\usepackage[document]{ragged2e} 							
\usepackage{color,soul}

\usepackage{graphicx}
\graphicspath{{../Introduction Figures/}{../Figs/}}	
\usepackage{tablefootnote}								
\usepackage[justification=raggedright,singlelinecheck=false,labelsep=period]{caption} 
\setkomafont{captionlabel}{\bfseries} 						
\setkomafont{caption}{\bfseries} 							
\setcapindent{0pt} 										
\usepackage{etoolbox}									
\AtBeginEnvironment{figure}{\setlength{\RaggedRightParindent}{0em}}
\AtBeginEnvironment{table}{\setlength{\RaggedRightParindent}{0em}}

\usepackage{chngcntr}									
\counterwithout{figure}{chapter}
\counterwithout{table}{chapter}

\usepackage[square,sort,comma,numbers,sort&compress]{natbib} 
\makeatletter
\renewcommand\NAT@bibsetnum[1]{\settowidth\labelwidth{\@biblabel{#1}}%
   \setlength{\leftmargin}{\bibindent}\addtolength{\leftmargin}{\dimexpr\labelwidth+\labelsep\relax}%
   \setlength{\itemindent}{-\bibindent}%
   \setlength{\listparindent}{\itemindent}
\setlength{\itemsep}{\bibsep}\setlength{\parsep}{\z@}%
   \ifNAT@openbib
     \addtolength{\leftmargin}{\bibindent}%
     \setlength{\itemindent}{-\bibindent}%
     \setlength{\listparindent}{\itemindent}%
     \setlength{\parsep}{0pt}%
   \fi
}
\makeatother

\usepackage[disable,textsize=tiny]{todonotes}   				

\usepackage{miller}  										

\usepackage{chemfig}
\usepackage{chemformula}								
\usepackage[colorlinks,citecolor=black,linkcolor=black,urlcolor=black]{hyperref}

\usepackage{setspace} 									
  {\begin{newspacing}{1}}{\end{newspacing}}
\newenvironment{newspacing}[1]%
  {\par\begingroup\newlinestretch{#1}}						
  {\par\vskip\parskip\vskip\baselineskip\endgroup
   \vskip-\parskip\vskip-\baselineskip} 
\newcommand{\newlinestretch}[1]							
  {\renewcommand{\baselinestretch}{#1}\currenttextsize}
\let\currenttextsize=\normalsize
  
\makeatletter
\renewcommand\sectionlinesformat[4]{%
  \@hangfrom{\hskip#2 #3}{#4} 								
}
\renewcommand\chapterlinesformat[3]{%
  \@hangfrom{#2}{#3}
}
\makeatother
\RedeclareSectionCommand[innerskip=0pt]{chapter} 			

\RedeclareSectionCommand[
  beforeskip=0pt,
  afterskip=0pt]{chapter}
\setkomafont{disposition}{\bfseries\normalsize}
\setkomafont{chapter}{\normalsize}

\setkomafont{section}{\normalsize} 
\RedeclareSectionCommand[
  beforeskip=0pt,
  afterskip=0.01pt]{section}
  
\setkomafont{subsection}{\normalsize} 
\RedeclareSectionCommand[
  beforeskip=0pt,
  afterskip=0.01pt]{subsection}

\KOMAoptions{toc=chapterentrydotfill} 						
\setkomafont{chapterentry}{} 								
\addtokomafont{chapterentrypagenumber}{\mdseries} 			
\addto\captionsenglish{} 

\RedeclareSectionCommand[tocnumwidth=3em]{chapter}			

\input{myConfig}

\input{glossary.tex}										

\begin{document}

\frontmatter
	
\include{title}

\include{signature}
\include{abstract}
\include{acknowledgements}

\clearpage
\tableofcontents

\clearpage \addcontentsline{toc}{chapter}{LIST OF TABLES}
\listoftables

\clearpage \addcontentsline{toc}{chapter}{LIST OF FIGURES}
\listoffigures

\mainmatter
\include{manuscript/chapter01}
\include{manuscript/chapter02}
\include{manuscript/chapter03}

\include{manuscript/chapter04}
\include{manuscript/chapter05}


\clearpage \phantomsection	
\addcontentsline{toc}{chapter}{REFERENCES} \renewcommand\bibname{REFERENCES \vspace{10 mm}}
\bibliographystyle{unsrt}
\bibliography{citations.bib} 

\end{document}

%% file: myConfig.tex
\def \cAuthorName {Abigail Kwan}
\def \cGraduationDate {Spring 2020 }

\def \cThesisTitle {Malware Detection at the Microarchitecture Level using Machine Learning Techniques}

\graphicspath{{../Figs/}{../Figs-generated/}}

\usepackage{gensymb}									
\usepackage{siunitx}										
\DeclareSIUnit\MU{m.u.}
\DeclareSIUnit\oersted{Oe}
\DeclareSIUnit\emu{emu}
\DeclareSIUnit\torr{torr}
\DeclareSIUnit\bohr{\mu_\text{B}}
\DeclareSIUnit\fu{\textit{f.u.}}

%% file: glossary.tex
\usepackage[symbols,nogroupskip]{glossaries-extra}
\glsxtrnewsymbol[description={d-spacing of lattice}]{d_b}{\ensuremath{d_b}}
\glsxtrnewsymbol[description={MnPc thickness determined from QCM}]{t_{QCM}}{\ensuremath{t_{QCM}}}

\glsxtrnewsymbol[description={positive saturation magnetization}]{M_{sat+}}{\ensuremath{M_{sat+}}}
\glsxtrnewsymbol[description={negative saturation magnetization}]{M_{sat-}}{\ensuremath{M_{sat-}}}
\glsxtrnewsymbol[description={average saturation magnetization}]{M_{sat}}{\ensuremath{M_{sat}}}
\glsxtrnewsymbol[description={average saturation magnetization per volume}]{m_{sat}}{\ensuremath{m_{sat}}}
\glsxtrnewsymbol[description={average saturation magnetization per formula unit and Bohr magneton}]{mu_{sat}}{\ensuremath{\mu_{sat}}}

\glsxtrnewsymbol[description={substrate deposition temperature}]{T_{dep}}{\ensuremath{T_{dep}}}
\glsxtrnewsymbol[description={sample measurement temperature}]{T_{m}}{\ensuremath{T_{m}}}
\glsxtrnewsymbol[description={lattice parameter of rhombohedral structure}]{a}{\ensuremath{a}}
\glsxtrnewsymbol[description={lattice parameter of monoclinic structure}]{c}{\ensuremath{c}}
\glsxtrnewsymbol[description={surface area of MnPc thin film}]{A}{\ensuremath{A}}
\glsxtrnewsymbol[description={volume of MnPc thin film}]{V}{\ensuremath{V}}

\glsxtrnewsymbol[description={base pressure of thermal evaporator}]{P_b}{\ensuremath{P_b}}
\glsxtrnewsymbol[description={Gaussian peak prominence (over background)}]{P_r}{\ensuremath{P_r}}

\glsxtrnewsymbol[description={full width half maximum in radians}]{beta}{\ensuremath{\beta}}
\glsxtrnewsymbol[description={X-ray radiation wavelength}]{lambda}{\ensuremath{\lambda}}

\glsxtrnewsymbol[description={Gaussian peak width}]{sigma}{\ensuremath{\sigma}}
\glsxtrnewsymbol[description={Debye-Scherrer crystalline size}]{tau}{\ensuremath{\tau}}
\glsxtrnewsymbol[description={DC susceptibility}]{chi}{\ensuremath{\chi}}

%% file: title.tex
\begin{titlepage}
    \begin{center}
        \linespread{2}
        \normalsize
        \textbf{ \cThesisTitle }\\
        \medskip
    \end{center}
    \begin{center}
        \linespread{1}
        \vfill

        		A THESIS \\
		Presented to the Department of Computer Engineering and Computer Science\\
        		California State University, Long Beach
        \vfill
        
        In Partial Fulfillment\\
        of the Requirements for \\
        the University Honors Program Certificate\\
        \vfill

        By \cAuthorName\\
        \bigskip

        \bigskip
		\cGraduationDate
    \end{center}
\end{titlepage}

%% file: abstract.tex
\chapter{\normalfont ABSTRACT}

\begin{center}

  \textbf{\cThesisTitle}\\

  By\\
  \cAuthorName \\
  \cGraduationDate

\end{center}

Detection of malware cyber-attacks at the processor microarchitecture level has
recently emerged as a promising solution to enhance the security of
computer systems. Security mechanisms, such as hardware-based malware detection, use machine learning algorithms to classify and detect malware with the aid of Hardware Performance Counters (HPCs)
information. The ML classifiers are fed microarchitectural data extracted from Hardware Performance Counters (HPCs), which contain behavioral data about a software program. These HPCs are captured at run-time to model the program's behavior. Since the amount of HPCs are limited per processor, many techniques employ feature reduction to reduce the amount of HPCs down to the most essential attributes. Previous studies have already used binary classification to implement their malware detection after doing extensive feature reduction. This results in a simple identification of software being either malware or benign. This research comprehensively analyzes different hardware-based malware detectors by comparing different machine learning algorithms' accuracy with binary and multi-class classification models. Our experimental results indicate that when compared to complex machine learning models (e. g. Neural Network and Logistic), light-weight J48 and JRip algorithms perform better in detecting the malicious patterns even with the introduction of multiple types of malware. Although their detection accuracy slightly lowers, their robustness (Area Under the Curve) is still high enough that they deliver a reasonable false positive rate. 

%% file: acknowledgements.tex
\chapter{Acknowledgements}

I would like to thank my mentor, Dr. Hossein Sayadi, for introducing me to this topic and guiding me throughout this process. I would also like to thank Mr. Kevin and Kent Peterson for the P2S Engineering Honors Scholarship. Lastly, I would like to thank my family and friends for supporting me through my college journey.

%% file: manuscript/chapter01.tex
\chapter{Introduction}
\par As the world is increasingly connected through the internet, many devices are prone to security threats and malware attacks. 
Malware  is  a  piece  of  code  designed  to  perform  various  malicious activities, such as destroying the data, stealing information, running destructive  or  intrusive  programs  on  devices  to  perform  Denial-of- Service (DoS) attacks, and gaining root access without the consent of user. These programs are used to compromise user data and cripple networks \cite{dinakarrao2019lightweight,sayadi2018customized}. The  recent  proliferation  of computing devices in mobile and Internet-of-Things (IoTs) domains further exacerbates  the  malware  attacks  and  calls  for  eﬀective  malware detection techniques \cite{sayadi_ensemble_2018,sayadi2018comprehensive}. A recent survey showed that the number of security incidents in 2014 rose to 42.8 million incidents \cite{ozsoy_hardware-based_2016}. Despite the number of antivirus software out on the market, malware still persist due to the broad number of virus variations that could exist. 

Conventional signature-based and semantic-based malware detection methods 
mostly impose significant computational overhead to the system. Furthermore, they are unable to detect unknown threats making them unsuitable for devices with limited available computing and memory resources \cite{sayadi_2smart:_2019}. The emergence of new malware threats requires patching or updating the software-based malware detection solutions (such as off-the-shelf anti-virus) that need a vast amount of memory and hardware resources, which is not feasible for emerging computing systems specially in embedded mobile and IoT devices \cite{sayadi2018customized}. 
A typical antivirus software uses static characteristics of malware to detect if it is a threat \cite{demme_feasibility_2013}. This is extremely exploitable as it requires an up-to-date database in which a new malware won't be detected \cite{damodaran_comparison_2017}. Since software antivirus programs can be bypassed, researchers have begun studying hardware-based malware detection in order to increase security for computers and networks.

Recent works
have demonstrated that malware can be differentiated from
normal applications by classifying anomalies using Machine
Learning (ML) techniques in low-level microarchitectural
feature spaces captured by Hardware Performance Counters
(HPCs) \cite{sayadi_ensemble_2018,demme_feasibility_2013,sayadi_2smart:_2019}. The HPCs are a set of special purpose registers built into modern microprocessors to capture
the trace of hardware-related events for a running program
 \cite{wang_malicious_2016}. ML-based malware detectors can be implemented
in microprocessor hardware with significantly low overhead as
compared to software-based methods, as detection inside
the hardware is very fast within few clock cycles \cite{sayadi_ensemble_2018, ozsoy_hardware-based_2016}.

\section{Related Work}
\par In this section, we discuss recent related work for hardware-based malware detection. The work in \cite{patel_analyzing_2017} had a similar approach when it came to data collection and classification. They also implemented the machine learning algorithms into software and hardware. They found that the software implementation had a large overhead, which caused the algorithm to lag behind due to latency. The hardware implementation was done using Vivado High-Level-Synthesis, which allowed them to collect data on power estimation, data, and latency. Their results found that more simple classifiers such as OneR were more efficient when implemented on hardware. Even though it was not as accurate, it was faster, took less power, and the least amount of area. 

\par Besides regular machine learning methods, some other studies have examined hardware-based detection using ensemble learning. Researchers in \cite{sayadi_ensemble_2018} used ensemble learning in their hardware implementation in order to find out how it could improve malware detection. The two techniques they used were boosting and bagging. Boosting is a technique that weighs each training dataset and adjusts the weights based on the overall accuracy of the model. Bagging, on the other hand, is another ensemble learning technique that takes a statistical value from multiple random samples and uses it to train the ML models.  They compared the robustness and the accuracy of regular classifiers with boosted classifiers. They found that boosting techniques improved the classification by as much as 17\% with a much lower amount of HPCs. 

\par  In addition, a recent work in \cite{sayadi_2smart:_2019} proposed a two-stage machine learning-based approach for run-time malware detection in which the first level classifies applications using a multiclass classification technique into either benign or one of the malware classes (Virus, Rootkit,
Backdoor, and Trojan). In the second level, to have a high detection performance, the authors deploy an ML model
that works best for each class of malware and further apply effective ensemble learning to enhance the performance of hardware-based malware detection. The work in \cite{sayadi2018customized} also proposed an effective machine learning-based hardware-assisted malware detection framework for embedded devices which only utilizes a limited number (only 4) of low-level features in a microprocessor i.e., HPC events to facilitate the run-time malware detection. 

\par Machine learning and ensemble learning are not the only methods to implement hardware-based malware detection. The researchers at \cite{alizadeh_akoman:_2018} used Akoman, a malware detection technique that builds behavioral signatures for malware detection. The technique collects signatures and creates an aggregation matrix. After that, it applies fast and exact signatures to compare the program to likely malware families in order to determine whether or not it is actually benign. They tested Akoman using 36 benign Linux programs and 13 families of Linux malware programs. They found that Akoman can achieve an acceptable performance based on its metrics for average precision, recall, and F-measure.

\par Recent research have also dealt with the theoretical side of hardware-based malware detection. The study in \cite{basu_theoretical_2020} analyzed the security guarantees of hardware-based malware detection using 4 HPCs. They calculated the probability of malware being detected when HPCs are monitored at a certain sampling interval. They used control-flow graphs to outline programs. Their research found that it is difficult for malware to match all possible HPCs, with the probability of matching four HPC parameters is $\exp^{-40}$. This means that HPCs are highly secure compared to normal antivirus software since malware would have a difficult time guessing the right HPCs to pass undetected. 

\section{Motivation for Research}
\par 
Hardware-based malware detection has become an increasingly important topic in the field of cybersecurity. Implementation at a hardware level reduces the chances of malware subverting protection mechanisms\cite{demme_feasibility_2013,sayadi_ensemble_2018}.Previous research has already determined classification of several malware types using memory access operations\cite{banin_multinomial_2018}. The researchers were able to classify various malware types based on the number of memory accesses. There also exists research that deal with binary classification of malware using hardware performance counters (HPC) \cite{patel_analyzing_2017,sayadi_ensemble_2018}. According to their findings, simple decision tree algorithms such as OneR were more effective in hardware implementation compared to higher accuracy algorithms such as logistics and MultiLayerPerceptron. Simple classifiers were found to be more cost-effective when it came to execution time, accuracy per logic area, and number of HPC used. This thesis will expand on these findings to determine if certain machine learning methods are more efficient with differentiating between multiple types of malware. Instead of only binary classification, this thesis will explore how certain machine learning methods perform in detecting five different types of malware: backdoor, rootkit, trojans, viruses, and worms. 

The thesis is organized as follows. Background information concerning malware, HPCs, and machine learning algorithms are detailed in \autoref{chap:2}. The experimental setup about the dataset and approach used is described in \autoref{chap:3}. In \autoref{chap:4}, we discuss the results collected and provide an analysis of the differences between binary and multiclass classification amongst the various ML classifiers. Finally, in \autoref{chap:5}, the conclusion and future work is presented for this thesis.  

%% file: manuscript/chapter02.tex
\chapter{Background}
\label{chap:2}
\section{Types of Malware}
\par{}
Malware can have multiple infection vectors. One way is through exploiting vulnerable services over a network\cite{egele_survey_2012}. Large scale installations of systems with the same vulnerability can allow malicious software to infect automatically. Another way malware can infect systems is through drive-by downloads\cite{egele_survey_2012}. The malware downloads automatically by exploiting a web browser vulnerability. Because of the variety in infection vectors and methods, malware can be categorized based on the purposes and ways they are able to reach vulnerable systems\cite{ye_survey_2017}. For the purposes of this thesis, this section gives an overview of the five different malware types that we are testing. 
\begin{itemize}
    \item {\textbf{Backdoor}}
    Backdoors are methods used to bypass regular authentication or encryption in a system. Backdoors are used alongside Trojans so attackers can have access to a remote computer or network. 
    \item {\textbf{Rootkit}}
    Rootkits are designed to be stealthy software that hide certain processes from a computer system. \cite{ye_survey_2017, egele_survey_2012}. 
    \item {\textbf{Trojans:}}
    Trojans are software programs that perform malicious attacks under the guise of being a regular program. \cite{ye_survey_2017}
    \item {{\textbf{Viruses:}}}
    Viruses are dependent malware that can attach to other system programs. When it is executed, the affected area gets infected.[ye] It spreads because it can infect not only local files, but also files on shared servers, thereby affecting other computers as well \cite{egele_survey_2012}.
    \item {\textbf{Worms:}}
    Worms are programs that are able to run independently. They propagate by copying itself from an infected host into another machine, usually through the operating system. \cite{ye_survey_2017}
\end{itemize}
\par{}
Because of the numerous types of malware,  a diversified dataset is needed in order to design an efficient malware detection system. Features must be collected from hardware performance counters (HPCs) using data mining. 

\section{Hardware Performance Counters}
Hardware performance counters are registers built into a microprocessor in order to store data about hardware events \cite{patel_analyzing_2017, wang_malicious_2016,sayadi_ensemble_2018}. These registers give information about the run-time behavior of software programs. The advantage of using HPCs is increased security. Unlike antivirus software which characterize malware based on static characteristics, HPCs characterize based on a program's actual behavior\cite{wang_hardware_2016}. HPCs are also not accessible to malware. The number of HPCs that are available on a processor vary depending on the model\cite{wang_malicious_2016}. For example, an Intel Atom processor only has 4 HPCs to monitor micro-architectural behavior at run-time \cite{patel_analyzing_2017}. Because of this limitation, it's important to narrow down the number of features so that the machine learning model can be implemented on hardware more efficiently to detect the malware. 

\section{Machine Learning Methods}
In a supervised learning process, a training set is used to create a classification model \cite{sayadi2017machine}. The learning algorithm is given a set of N samples with A attributes such that when given new data, it accurately predicts the class\cite{baranauskas_tree-based_2018}. The training examples are tuples (x, y), where x refers to the sample and y refers to the class\cite{baranauskas_tree-based_2018}. In this thesis, the data will be put through five machine learning algorithms: Multilayer Perceptron (MLP), OneR, Logistic, JRip, and J48. 
\begin{figure}
\centering
\includegraphics[scale=0.8]{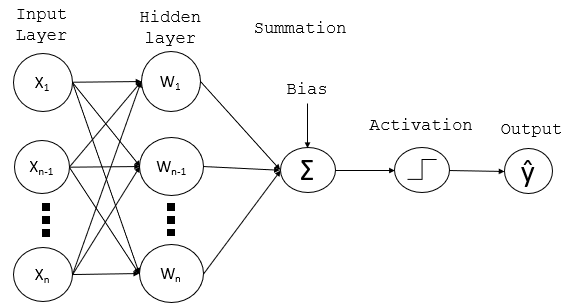}
\caption{The Multilayer Perceptron neural network has hidden layers that are not directly exposed to the input. This particular figure is a feed-forward network. Its purpose is to approximate some function f(x). }
\label{fig:MLPProcessDiagram}
\end{figure}
\begin{description}
\item[Multilayer Perceptron]: The Multilayer Perceptron (MLP) (Figure \ref{fig:MLPProcessDiagram}) algorithm is a neural network that makes predictions based on mapping. It contains multiple layers with one or more hidden layers \cite{pano-azucena_fpga-based_2018}.  MLP is a type of logistic regression where the input is transformed based on weighting values. The weights on each node can be changed after data is processed and the error is calculated. After the weights are calculated, the weights are plugged into an activation function to approximate the output. The MLP function is defined as\cite{trenn_multilayer_2008}: 
\begin{equation}
    f_{MLP}: R_{n0} \xrightarrow[]{} R, x = (x_1, x_2, ... x_{n0}) \xrightarrow[]{} y
\end{equation} Where x refers to the number of inputs while R refers to the activation function. The standard activation function used in most MLPs is $\sigma (t) = 1/(1+e^{-t})$ \cite{trenn_multilayer_2008}. The activation function is used after the hidden layers have been calculated. 

\item[OneR]:
The OneR is a simple decision tree based algorithm. It builds one rule for each attribute in the training set, then selects the rule with the smallest error\cite{deeba_classification_2016}. The rule is built by finding out the most frequent class, which is simply the class that appears the most for that value. The algorithm is as follows\cite{nevill-manning_development_1995}: 

\begin{verbatim}
1 For each attribute x, form a rule: 
2   For each value y from the domain of x, 
3    Select the set of instances where y has value x, 
4       Let c be the most frequent class in that set. 
5       Add the following clause to the rule for x: 
6           if x has value y then the class is c 
7   Calculate the classification accuracy of this rule 
8 Use the rule with the highest classification accuracy. 
\end{verbatim}
\item[Logistic Regression]:
Logistic Regression is a linear classifier that uses assumes the outcome is a linear function of independent variables. The probability of the linear combination in logistic regression can be represented with the following formula:\cite{ngufor_extreme_2016} 
\begin{equation}
    \pi(x) = Pr(y = 1|x) 
\end{equation}
This formula can be further expanded by defining it as: 
\begin{equation}
    \pi(x) = Pr(y = 1|x;\beta) = 1/(1+\exp(-\beta_0 - \beta^Tx)
\end{equation}
$\beta_0$ is a constant that refers to the bias in the equation. $\beta^Tx$ refers to unknown values in the model. Therefore, logistic regression models the probability of a variable y taking on the value y = 1 depending on a number of independent variables x. 

\item[JRip]:
JRip is a rule-based classifier which creates propositional rules that can be used to classify elements\cite{tarun_generating_2014}. It takes the instances in the dataset and evaluates them in increasing order \cite{bhargava_detection_2018}. JRip then generates a set of rules and attributes for that particular class before moving on to the next class.
\item[J48]:
J48 is a lightweight classifier that creates a decision tree using the C4.5 algorithm. The C4.5 algorithm is a decision tree algorithm that uses gain ratio to select the best "splitting" feature when making a decision tree\cite{li_effective_2018}. Gain Ratio is defined as 
\begin{equation}
    GainRatio(A) = \frac{Gain(A)}{SplitInfo(A)}, 
\end{equation}
where Gain(A) represents Information Gain for feature A and SplitInfo(A) represents the potential information generated by splitting the set D over n outcomes according to feature A. SplitInfo can be further defined as 
\begin{equation}
    SplitInfo_{A}(\mathcal {D}) = -\sum _{j=1}^{k} \frac {| \mathcal {D}_{j} |}{| \mathcal {D} |} \times \log _{2}\left({\frac {| \mathcal {D}_{j} |}{| \mathcal {D} |}}\right).
\end{equation}
\end{description}

\section{Binary vs. Multiclass Classification} 
Binary classification is the simplest type of machine problem that takes an element and classifies it into one of two groups. It determines the grouping of an element by looking at its attributes. In the context of malware detection, binary classification determines whether a program is malware or benign. What makes multiclass classification different from binary classification is the number of classes that must be taken into account. The set of target classes can grow dynamically and is not fixed, which can make it increasingly complex \cite{meng_joo_er_online_2016}. For malware detection, it means that the machine learning algorithm must identify if a software is a specific type of malware. 

%% file: manuscript/chapter03.tex
\chapter{Experimental Setup}
\label{chap:3}
\section{Dataset}
\par{}
For training and testing the machine learning models we used a  per-class dataset of microarchitectural features in \cite{sayadi_2smart:_2019}. The applications  were executed on an Intel Xeon X5550 machine
running Ubuntu 14.04 with Linux 4.4 Kernel. The dataset contains complete samples collected from HPCs for both benign and malware. The malware data can be further divided into five categories. There are 16 features collected for this dataset. Some of the features included are bus cycles, cache misses, branch instructions, etc. The dataset was collected using Perf, which is a tool available on the Linux operating system. Perf can measure multiple events simultaneously \cite{sayadi_2smart:_2019}. Perf collects the data based on the execution of both benign and malware applications. The programs are run in isolated environments called Linux Containers which provide access to actual performance instead of emulated HPCs\cite{sayadi_2smart:_2019}. A total of about 12000 samples were taken from the dataset and combined in Excel to achieve a 70\% - 30\% data split. For binary classification, the data was combined to ensure about 70\% malware - 70\% benign for the training set and 30\% malware - 30\% benign for the test set. For multiclass classification, the same number of samples was used except the malware data was replaced with more specific classifiers. Instead of just malware, it would be labeled trojan, rootkit, etc. 

\section{WEKA} 
\par
\begin{figure}
\centering
\includegraphics[scale=0.5]{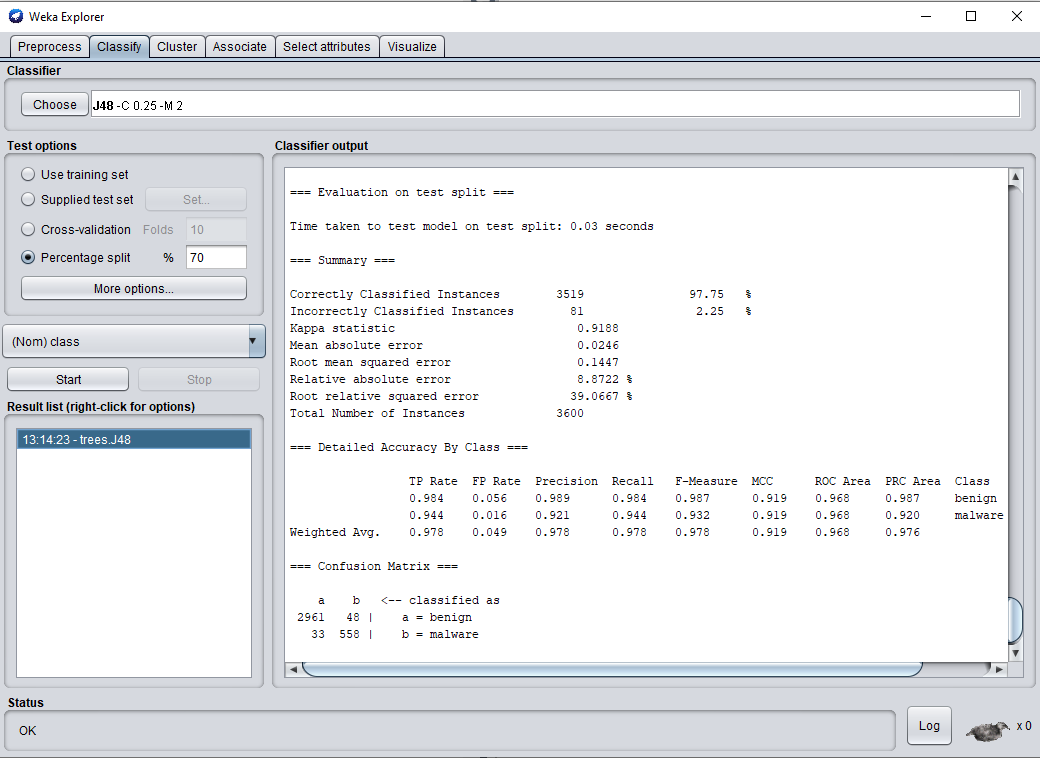}
\caption{WEKA's Explorer application contains several tools that can be used for applying machine learning. The Classify tab is where the algorithms are applied to create a model that can be used for test sets.}
\label{fig:WEKASnapshotWindow}
\end{figure}

{}
For this thesis, we are using WEKA to process the dataset and obtain the results for classification accuracy. WEKA is a software tool developed by researchers at the University of Waikato that contains a collection of machine learning algorithms for data mining. For this experiment, we are using the \emph{Explorer} application of WEKA for both feature selection and classification. The \emph{Explorer} application has several tabs that aid in processing the dataset. For feature selection, the \emph{Select Attributes} tab is used. This tab contains the Attribute Evaluator which can be set by the user to select different types of attribute evaluation methods. It also has a section where the search method can be selected. For classification, the machine learning algorithms are selected under the \emph{Classify} tab. In this tab, the algorithm can be selected under the Classifier section. This tab also allows for customization for testing under the test options section. The results of the test set can be seen in the Classifier output window. An example of how the \emph{Classify} tab looks can be seen in Figure \ref{fig:WEKASnapshotWindow}. 

\section{Feature Selection}
\begin{figure}
\centering
\includegraphics[scale=0.7]{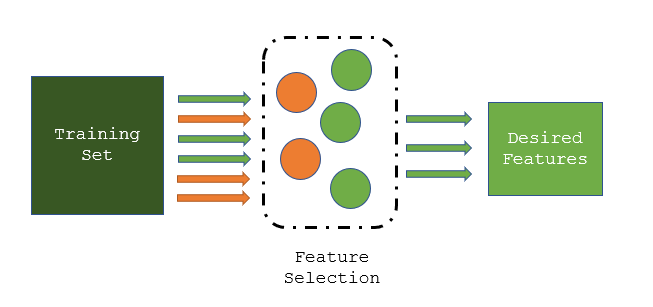}
\caption{The feature selection process selects relevant predictors from a training set in order to increase the accuracy of the prediction variable. In this model, orange represents redundant or impure classifiers while green represents the desired classifier.}
\label{fig:FeatureSelectionDiagram}
\end{figure}

Feature selection must be used on the data in order to find the best possible classifiers to use with the limited amount of HPCs \cite{sayadi_ensemble_2018,sayadi_2smart:_2019} (Figure ~\ref{fig:FeatureSelectionDiagram}). Previous studies have indicated that many algorithms do poorly with large numbers of irrelevant features\cite{baranauskas_tree-based_2018}. Performing feature selection on the data can also increase the accuracy of the learning algorithm. 
\par{} For the purposes of this thesis, the feature selection algorithm that will be used is Information Gain. Information Gain makes use of entropy, which is a common quantity in information theory associated with any random variable. Since entropy measures impurities in the training set, we can define Information Gain as the entropy of the training set minus the average entropy of the child nodes. Information Gain is represented using the formula\cite{lippi_relational_2011}: 
\begin{equation}
    IG(X, y) = H(X) - H(X|y)  
\end{equation}
IG(X, y) refers to the information in the dataset for a specific random variable y. H(X) refers to the entropy for the training set, while H(X|y) refers to the conditional entropy based on the random variable y. 
After the data is collected, the extracted data must go through pre-processing. The data is subdivided into 70\% - 30\% split with 70\% for the training set and 30\% for the test set. Since HPCs only have a limited amount of registers, the number of attributes per data must be reduced. Using WEKA's Attribute Evaluator, features are reduced using Information Gain Attribute Evaluation. The search method used is Ranker. Features are then reduced by taking the top 8 features from the results of the training set, then further reduced down up until it reaches 2 features.
\par{}
For the purpose of this thesis, we used feature selection for both binary and multi-classification. As seen in Table \ref{tab:binclas_order} and Table \ref{tab:multclas_order}, the top 8 features for both types of classification is very similar. The order only differs slightly, which tells us that these top features are the ones that matter most for malware detection for this specific dataset. L1-dcache-loads and L1-dcache-stores are the top 2 HPCs after extensive feature selection, which means that these two HPCs are really important to use when implementing the actual machine learning onto hardware. 

\begin{table}[]
    \centering
    \includegraphics[scale=0.7]{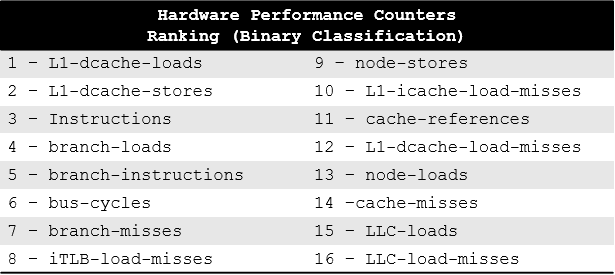}
    \caption{Binary Classification Order of Attributes}
    \label{tab:binclas_order}
\end{table}
\begin{table}[]
    \centering
    \includegraphics[scale=0.7]{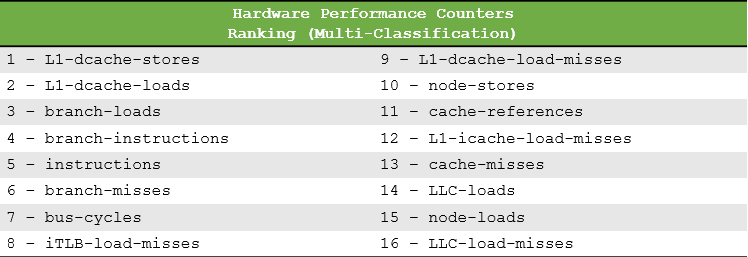}
    \caption{Multiclass Classification Order of Attributes}
    \label{tab:multclas_order}
\end{table}

\newpage
\section{Classification}
\begin{figure}
    \centering
    \includegraphics[scale=0.6]{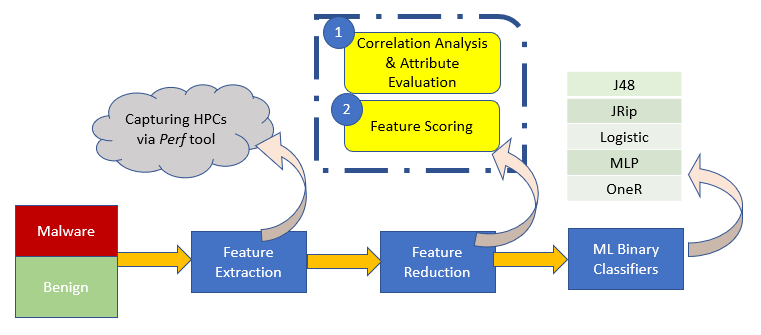}
    \caption{Overview of run-time hardware-based malware detection approach \cite{sayadi_ensemble_2018}}
    \label{fig:detectionapproach}
\end{figure}

Once the desired features have been determined, the training set must go through WEKA's Classifier. For each classifier, we supply the test set. WEKA then details the accuracy of each classifier in the output section. It also provides a confusion matrix that details how many were classified correctly for each attribute. The results also detail important information such as ROC Area and false positive rate. After obtaining the results, we graph the ROC Area by visualizing the threshold curve and extracting the true positive and false positive rates. The general approach for the classification was based on a previous study\cite{sayadi_ensemble_2018}. An overview can be seen in the Figure \ref{fig:detectionapproach}. 

%% file: manuscript/chapter04.tex
\chapter{Results and Analysis}
\label{chap:4}
\section{Accuracy of Classification}

\begin{figure}[]
    \centering
    \subfloat[Binary Classification]{{ \includegraphics[scale=0.8]{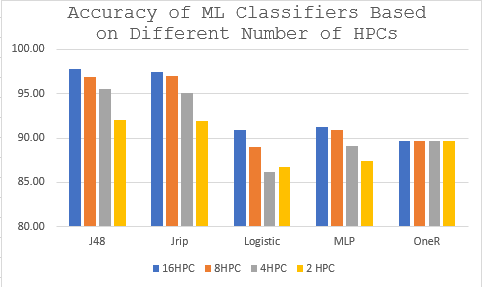} }}%
    \qquad
    \subfloat[Multiclass Classification]{{\includegraphics[scale=0.8]{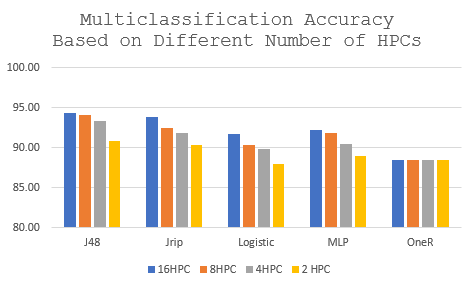}}}%
    \caption{Accuracy of Classification Models per Number of HPCs}%
    \label{fig:accuracy_ROC}%
\end{figure}

\par{}
To evaluate the efficiency of the number of HPCs used, the percentage accuracy of each classification was calculated. The accuracy was calculated for different numbers of HPCs (16, 8, 4, and 2). As seen in Figure \ref{fig:accuracy_ROC}, the accuracy of the classifiers tend to go down as the number of HPCs decreased. More complex algorithms such as Logistic and MLP had the lowest percentage accuracy. OneR had no discernible decrease because the algorithm only relies on one HPC to predict the behavior. For the multiclass classification model, the accuracy took a hit for all five classifiers. Since the same amount of malware samples were used for both binary and multiclass classification, having more specific classes reduced the accuracy of malware identification. Figure \ref{fig:accuracy_ROC} shows how the accuracy is much lower for all classifiers. Noticeably, the more complex logistic and MLP did better than the OneR when it came to classifying different types of malware.  

\begin{table}[]
    \centering
    \includegraphics[scale=0.7]{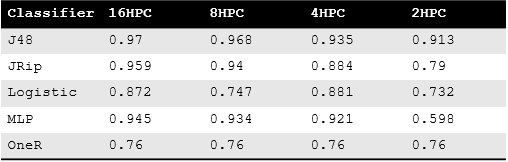}
    \caption{Binary Classification ROC Area for Different Number of HPCs}
    \label{tab:binclas_roctable}
\end{table}

\section{Comparison of Receiver Operating Characteristics} 
\par
\begin{figure}[]
    \centering
    \subfloat[8 HPCs]{{ \includegraphics[scale=0.8]{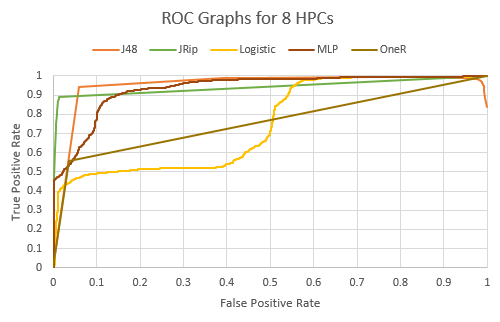} }}%
    \qquad
    \subfloat[4 HPCs]{{\includegraphics[scale=0.8]{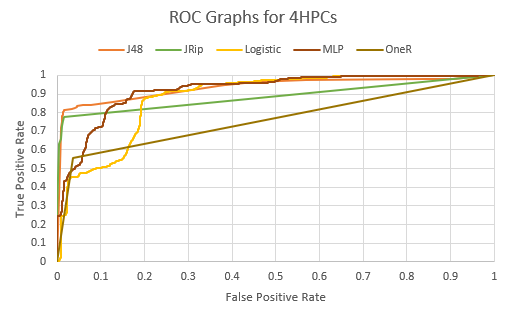}}}%
    \caption{Binary Classification ROC Graphs with Varying Number of HPCs}%
    \label{fig:binaryclass_ROC}%
\end{figure}

{}Another metric to observe is the Receiving Operating Characteristics (ROC) result. ROC measures the Area Under the Curve (AUC) for evaluating the robustness of each ML classifier\cite{sayadi_ensemble_2018,sayadi2018comprehensive}. The ROC corresponds to the probability that the classifier correctly identified which application is malware and which is benign. Using this data, it is possible to find the threshold where the classifier achieves 100\% accurate positive identification. As seen in Table \ref{tab:binclas_roctable}, the ROC Area results correspond to the accuracy of the ML classifiers. A notable result is the sharp drop for the MLP classifier when it goes from 4 HPC down to 2HPC. J48 also dropped but didn't lose as much ROC compared to the other classifiers. This result tells us that having 4 HPCs is an ideal amount of HPCs without losing too much accuracy for binary classification. The visualization of the ROC area can be seen in Figure \ref{fig:binaryclass_ROC} for 8HPCs and 4HPCs. From these graphs, we can see that the reduction from 8 HPCs down to 4 HPCs reduces the true positive rate of most ML classifiers. It also shows, however, that some classifiers do well better than others even with feature reduction. For example, JRip and J48 still did better than the other classifiers, with J48 maintaining an ROC Area over 0.9 for all different number of HPCs. This gives valuable insight on which ML Classifier is best depending on the number of HPCs available. 

\begin{table}[]
    \includegraphics[scale = 0.8]{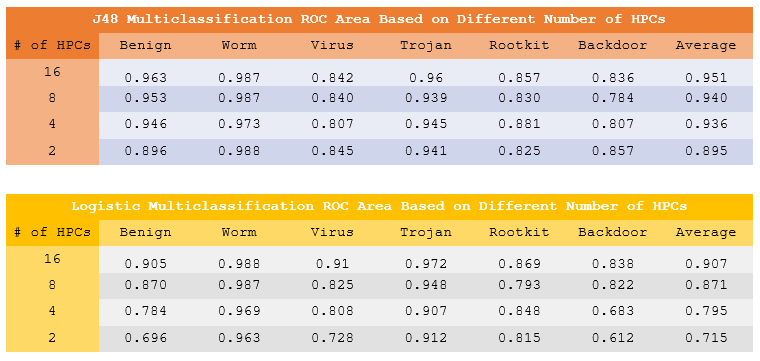}
    \caption{J48 and Logistic Multiclass Classification Models' ROC Area for Varying Numbers of HPCs}
    \label{tab:j48log_rocarea}
\end{table}

\par
\begin{figure}[]
    \centering
    \subfloat[Logistic]{{ \includegraphics[scale=0.8]{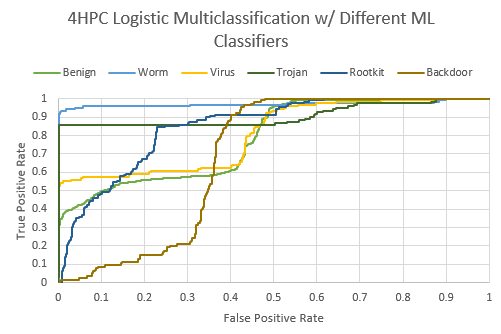} }}%
    \qquad
    \subfloat[J48]{{\includegraphics[scale=0.8]{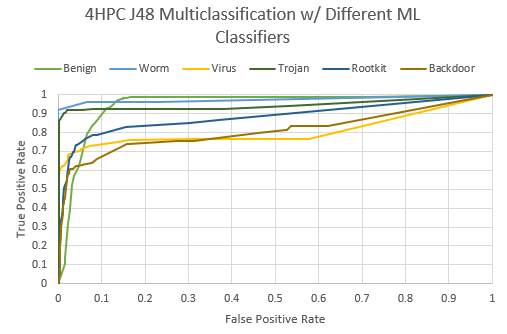}}}%
    \caption{Multiclass Classification 4HPCs ROC Graphs}%
    \label{fig:multiclass_ROC}%
\end{figure}

{}
Unlike binary classification, the ROC area in multiclass classification models vary depending on the type of malware class. For this thesis, we show the multiclass classification results for two machine learning algorithms: J48 and Logistic (Table \ref{tab:j48log_rocarea}). J48 outperformed the rest of the classifiers, on average. It did worse than logistic at identifying certain malware when the number of HPCs were high, but it did better when the number of HPCs were low. The ROC Area for 4HPCS of both J48 and logistic are represented as AUC graphs in Figure \ref{fig:multiclass_ROC}. As observed, the logistic multiclass classification model really suffers in performance when there are a low amount of HPCS. The J48 model, on the other hand, seems to have consistent performance for all types of malware. 

\par{}
From this we can conclude that the complexity of an algorithm has no advantage on accurately detecting multiple classes of malware. Even though logistic was more accurate at 8-16 HPCs, that scenario is usually unlikely to be implemented at a hardware level due to the limitations in the number of HPCs per processor. Therefore, implementing lightweight algorithms such as J48 may be more efficient when implementing multi-classification in the hardware level. 

%% file: manuscript/chapter05.tex
\chapter{Conclusion}
\label{chap:5}
Hardware-based malware detection is an emerging 
field in cybersecurity that implements machine learning-based malware detectors onto microprocessors. Hardware-based detectors rely on machine learning classiﬁers and use microarchitectural events captured by Hardware Performance Counters (HPCs) at run-time to identify the malicious patterns. Recent studies have shown that hardware performance and detection accuracy highly depend
on the type of machine learning used, as well as the type of HPCs used. Previous works have used binary classification to test the implementation of hardware-based malware detection. In this thesis, we compared and contrasted the results of binary and multiclass classification models to find out the differences when it comes to accuracy and detection performance. In our results, we found that the multiclass classification models depended on similar features as the binary classification models. We also found that out of the five machine learning algorithms used, the most efficient methods were J48 and JRip, which both performed well even with the introduction of multiple types of malware. The robustness of the J48 algorithm was also proven further with the ROC graphs for both binary and multiclass classification. Given the variety of current malware, it is more efficient to implement this type of algorithm for hardware-based detection in order to better detect different malware. 
\section{Future Work}
Currently, we are working on an ongoing research to implement multiclass classification machine learning detectors in MATLAB for hardware-based malware detection. MATLAB has a toolbox called HDL Coder that allows for the synthesis of MATLAB code into VHDL. Using HDL Coder as a tool for hardware-based malware detection is a new direction in this research as previous works have used Vivado High-Level Synthesis instead. 
\par{} Currently, I am working with the OneR algorithm as a case study for testing the conversion process of HDL Coder. I implemented the OneR algorithm in MATLAB by following Holte's 1R Classifier \cite{nevill-manning_development_1995}. The current MATLAB code takes in a set of vectors as a training set and outputs the top feature after calculating the accuracy of each feature. Data is first sorted into temporary vectors based on their class (malware or benign). The values in these vectors are then quantisized to find the right rules for each feature. After that, the values are tested for each feature using the rules that were created from quantisization. To calculate the accuracy, it first counts how many values were predicted correctly compared to their original class in the training set. The code then takes that total and divides it by the original number of data in the training set. The resulting percentage is the accuracy of that feature. We are working towards proposing the hardware implementations of binary and multiclass classification models used for hardware-based malware detection and extracting the hardware overhead and resource utilization characteristics of the detectors such area, latency, and power consumption. 

%% file: thesis.bbl
\begin{thebibliography}{10}

\bibitem{sayadi_ensemble_2018}
Hossein Sayadi, Nisarg Patel, Sai~Manoj P.D., Avesta Sasan, Setareh Rafatirad,
  and Houman Homayoun.
\newblock Ensemble {Learning} for {Effective} {Run}-{Time} {Hardware}-{Based}
  {Malware} {Detection}: {A} {Comprehensive} {Analysis} and {Classification}.
\newblock In {\em 2018 55th {ACM}/{ESDA}/{IEEE} {Design} {Automation}
  {Conference} ({DAC})}, pages 1--6, San Francisco, CA, June 2018. IEEE.

\bibitem{dinakarrao2019lightweight}
Sai Manoj~Pudukotai Dinakarrao, Hossein Sayadi, Hosein~Mohammadi Makrani,
  Cameron Nowzari, Setareh Rafatirad, and Houman Homayoun.
\newblock Lightweight node-level malware detection and network-level malware
  confinement in iot networks.
\newblock In {\em 2019 Design, Automation \& Test in Europe Conference \&
  Exhibition (DATE)}, pages 776--781. IEEE, 2019.

\bibitem{sayadi2018customized}
Hossein Sayadi, Hosein~Mohammadi Makrani, Onkar Randive, Sai~Manoj PD, Setareh
  Rafatirad, and Houman Homayoun.
\newblock Customized machine learning-based hardware-assisted malware detection
  in embedded devices.
\newblock In {\em 2018 17th IEEE International Conference On Trust, Security
  And Privacy In Computing And Communications/12th IEEE International
  Conference On Big Data Science And Engineering (TrustCom/BigDataSE)}, pages
  1685--1688. IEEE, 2018.

\bibitem{sayadi2018comprehensive}
Hossein Sayadi, Sai~Manoj P~D, Amir Houmansadr, Setareh Rafatirad, and Houman
  Homayoun.
\newblock Comprehensive assessment of run-time hardware-supported malware
  detection using general and ensemble learning.
\newblock In {\em Proceedings of the 15th ACM International Conference on
  Computing Frontiers}, pages 212--215, 2018.

\bibitem{ozsoy_hardware-based_2016}
Meltem Ozsoy, Khaled~N. Khasawneh, Caleb Donovick, Iakov Gorelik, Nael
  Abu-Ghazaleh, and Dmitry Ponomarev.
\newblock Hardware-{Based} {Malware} {Detection} {Using} {Low}-{Level}
  {Architectural} {Features}.
\newblock {\em IEEE Transactions on Computers}, 65(11):3332--3344, November
  2016.

\bibitem{sayadi_2smart:_2019}
Hossein Sayadi, Hosein~Mohammadi Makrani, Sai~Manoj Pudukotai~Dinakarrao,
  Tinoosh Mohsenin, Avesta Sasan, Setareh Rafatirad, and Houman Homayoun.
\newblock {2SMaRT}: {A} {Two}-{Stage} {Machine} {Learning}-{Based} {Approach}
  for {Run}-{Time} {Specialized} {Hardware}-{Assisted} {Malware} {Detection}.
\newblock In {\em 2019 {Design}, {Automation} \& {Test} in {Europe}
  {Conference} \& {Exhibition} ({DATE})}, pages 728--733, Florence, Italy,
  March 2019. IEEE.

\bibitem{demme_feasibility_2013}
John Demme, Matthew Maycock, Jared Schmitz, Adrian Tang, Adam Waksman, Simha
  Sethumadhavan, and Salvatore Stolfo.
\newblock On the feasibility of online malware detection with performance
  counters.
\newblock In {\em Proceedings of the 40th {Annual} {International} {Symposium}
  on {Computer} {Architecture} - {ISCA} '13}, pages 559--570, Tel-Aviv, Israel,
  2013. ACM Press.

\bibitem{damodaran_comparison_2017}
Anusha Damodaran, Fabio~Di Troia, Corrado~Aaron Visaggio, Thomas~H. Austin, and
  Mark Stamp.
\newblock A comparison of static, dynamic, and hybrid analysis for malware
  detection.
\newblock {\em Journal of Computer Virology and Hacking Techniques},
  13(1):1--12, February 2017.

\bibitem{wang_malicious_2016}
Xueyang Wang, Charalambos Konstantinou, Michail Maniatakos, Ramesh Karri,
  Serena Lee, Patricia Robison, Paul Stergiou, and Steve Kim.
\newblock Malicious {Firmware} {Detection} with {Hardware} {Performance}
  {Counters}.
\newblock {\em IEEE Transactions on Multi-Scale Computing Systems},
  2(3):160--173, July 2016.

\bibitem{patel_analyzing_2017}
Nisarg Patel, Avesta Sasan, and Houman Homayoun.
\newblock Analyzing {Hardware} {Based} {Malware} {Detectors}.
\newblock In {\em Proceedings of the 54th {Annual} {Design} {Automation}
  {Conference} 2017 on - {DAC} '17}, pages 1--6, Austin, TX, USA, 2017. ACM
  Press.

\bibitem{alizadeh_akoman:_2018}
Niloofar~S. Alizadeh and Mahdi Abadi.
\newblock Akoman: {Hardware}-{Level} {Malware} {Detection} {Using} {Discrete}
  {Wavelet} {Transform}.
\newblock In {\em 2018 {IEEE} {International} {Conference} on {Smart}
  {Computing} ({SMARTCOMP})}, pages 476--481, Taormina, June 2018. IEEE.

\bibitem{basu_theoretical_2020}
Kanad Basu, Prashanth Krishnamurthy, Farshad Khorrami, and Ramesh Karri.
\newblock A {Theoretical} {Study} of {Hardware} {Performance}
  {Counters}-{Based} {Malware} {Detection}.
\newblock {\em IEEE Transactions on Information Forensics and Security},
  15:512--525, 2020.

\bibitem{banin_multinomial_2018}
Sergii Banin and Geir~Olav Dyrkolbotn.
\newblock Multinomial malware classification via low-level features.
\newblock {\em Digital Investigation}, 26:S107--S117, July 2018.

\bibitem{egele_survey_2012}
Manuel Egele, Theodoor Scholte, Engin Kirda, and Christopher Kruegel.
\newblock A survey on automated dynamic malware-analysis techniques and tools.
\newblock {\em ACM Computing Surveys}, 44(2):1--42, February 2012.

\bibitem{ye_survey_2017}
Yanfang Ye, Tao Li, Donald Adjeroh, and S.~Sitharama Iyengar.
\newblock A {Survey} on {Malware} {Detection} {Using} {Data} {Mining}
  {Techniques}.
\newblock {\em ACM Computing Surveys}, 50(3):1--40, October 2017.

\bibitem{wang_hardware_2016}
Xueyang Wang, Sek Chai, Michael Isnardi, Sehoon Lim, and Ramesh Karri.
\newblock Hardware {Performance} {Counter}-{Based} {Malware} {Identification}
  and {Detection} with {Adaptive} {Compressive} {Sensing}.
\newblock {\em ACM Transactions on Architecture and Code Optimization},
  13(1):1--23, April 2016.

\bibitem{sayadi2017machine}
Hossein Sayadi, Nisarg Patel, Avesta Sasan, and Houman Homayoun.
\newblock Machine learning-based approaches for energy-efficiency prediction
  and scheduling in composite cores architectures.
\newblock In {\em 2017 IEEE International Conference on Computer Design
  (ICCD)}, pages 129--136. IEEE, 2017.

\bibitem{baranauskas_tree-based_2018}
José~Augusto Baranauskas, Oscar~Picchi Netto, Sérgio~Ricardo Nozawa, and
  Alessandra~Alaniz Macedo.
\newblock A tree-based algorithm for attribute selection.
\newblock {\em Applied Intelligence}, 48(4):821--833, April 2018.

\bibitem{pano-azucena_fpga-based_2018}
Ana Pano-Azucena, Esteban Tlelo-Cuautle, Sheldon Tan, Brisbane Ovilla-Martinez,
  and Luis de~la Fraga.
\newblock {FPGA}-{Based} {Implementation} of a {Multilayer} {Perceptron}
  {Suitable} for {Chaotic} {Time} {Series} {Prediction}.
\newblock {\em Technologies}, 6(4):90, October 2018.

\bibitem{trenn_multilayer_2008}
S.~Trenn.
\newblock Multilayer {Perceptrons}: {Approximation} {Order} and {Necessary}
  {Number} of {Hidden} {Units}.
\newblock {\em IEEE Transactions on Neural Networks}, 19(5):836--844, May 2008.

\bibitem{deeba_classification_2016}
K.~Deeba and B.~Amutha.
\newblock Classification {Algorithms} of {Data} {Mining}.
\newblock {\em Indian Journal of Science and Technology}, 9(39), October 2016.

\bibitem{nevill-manning_development_1995}
C.G. Nevill-Manning, G.~Holmes, and I.H. Witten.
\newblock The development of {Holte}'s {1R} classifier.
\newblock In {\em Proceedings 1995 {Second} {New} {Zealand} {International}
  {Two}-{Stream} {Conference} on {Artificial} {Neural} {Networks} and {Expert}
  {Systems}}, pages 239--242, Dunedin, New Zealand, 1995. IEEE Comput. Soc.
  Press.

\bibitem{ngufor_extreme_2016}
Che Ngufor and Janusz Wojtusiak.
\newblock Extreme logistic regression.
\newblock {\em Advances in Data Analysis and Classification}, 10(1):27--52,
  March 2016.

\bibitem{tarun_generating_2014}
Ivy~M. Tarun, Bobby~D. Gerardo, and Bartolome~T. Tanguilig~III.
\newblock Generating {Licensure} {Examination} {Performance} {Models} {Using}
  {PART} and {JRip} {Classifiers}: {A} {Data} {Mining} {Application} in
  {Education}.
\newblock {\em International Journal of Computer and Communication
  Engineering}, 3(3):202--207, 2014.

\bibitem{bhargava_detection_2018}
Neeraj Bhargava, Aakanksha Jain, Abhishek Kumar, and Dac-Nhuong Le.
\newblock Detection of {Malicious} {Executables} {Using} {Rule} {Based}
  {Classification} {Algorithms}.
\newblock pages 35--38, January 2018.

\bibitem{li_effective_2018}
Longjie Li, Yang Yu, Shenshen Bai, Ying Hou, and Xiaoyun Chen.
\newblock An {Effective} {Two}-{Step} {Intrusion} {Detection} {Approach}
  {Based} on {Binary} {Classification} and \$k\$ -{NN}.
\newblock {\em IEEE Access}, 6:12060--12073, 2018.

\bibitem{meng_joo_er_online_2016}
{Meng Joo Er}, Rajasekar Venkatesan, and {Ning Wang}.
\newblock An online universal classifier for binary, multi-class and
  multi-label classification.
\newblock In {\em 2016 {IEEE} {International} {Conference} on {Systems}, {Man},
  and {Cybernetics} ({SMC})}, pages 003701--003706, Budapest, Hungary, October
  2016. IEEE.

\bibitem{lippi_relational_2011}
Marco Lippi, Manfred Jaeger, Paolo Frasconi, and Andrea Passerini.
\newblock Relational information gain.
\newblock {\em Machine Learning}, 83(2):219--239, May 2011.

\end{thebibliography}
